# Analysis of C-point and V-point Inhomogeneous Vector Beams' Stability under Anisotropic Turbulence in Ground-to-Air Transmission

**Sagar[1,§], Sanjay Kumar Mishra[2] and Akhilesh Kumar Mishra[1,3,*]**

[1]*Centre for Photonics and Quantum Communication Technology, Indian Institute of Technology Roorkee, Roorkee-247667, Uttarakhand, India*

[2]*Adaptive Optics Group, Instruments Research and Development Establishment, Dehradun-248008, India*

[3]*Department of Physics, Indian Institute of Technology Roorkee, Roorkee-247667, Uttarakhand, India*

*Author e-mail address:* [§]sagar@cpqct.iitr.ac.in, [*]akhilesh.mishra@ph.iitr.ac.in

**Abstract:** In the present work, we explore the impact of anisotropic turbulence on the propagation characteristics of polarization singular LG and circular Airy vector beams along vertical, oblique, and horizontal transmission configurations. The influence of altitude-dependent turbulence anisotropy on beam evolution is assessed through changes in intensity distribution and degradation of spatial structure. Polarization characteristics of the propagated fields exhibiting C-point and V-point polarization singularities are quantified using the Stokes parameter framework, from which the degree of polarization is derived based on ensemble-averaged electric field correlations. To further interpret polarization dynamics, the corresponding Stokes vectors are mapped onto the higher order Poincaré sphere, providing a geometric visualization of turbulence-driven depolarization and polarization state dispersion. Numerical simulations have shown that the anisotropy of turbulence significantly enhances the polarization instability. Horizontal and slant-path propagations suffer stronger depolarization and polarization-state diffusion than vertical propagation. These results demonstrate the importance of vectorial polarization analysis in describing the propagation of a structured beam in real atmospheric environments. Overall, C-point polarization singularities of circular Airy vector beams are found to be the most resistant to turbulence-induced intensity fluctuation and polarization degradation in anisotropic atmospheric conditions.

## 1. Introduction

The investigation of optical beam propagation across air turbulence has been of significant interest in free space optical communication [1], imaging applications [2], and quantum communication [3-4]. Earlier studies have primarily concentrated on the propagation of scalar optical beams, particularly Gaussian [5] and Laguerre-Gaussian modes [6-8]. In contrast, recent progress in photonics has enabled the controlled generation and manipulation of vector-structured beams, which possess spatially inhomogeneous polarization states such as radial, azimuthal, and hybrid polarization configurations [9-11]. Vector beams have been shown to be more resistant to phase, amplitude, and scintillation index (SI) fluctuations caused by turbulence than scalar beams, due to their spatially variable polarization. [12]. Therefore, these beams can be regarded as promising candidates for optical information transfer of the future generation of communication.

The atmosphere causes random variations in the refractive index which distort the propagating beam parameters including phase, amplitude, polarization and spatial mode structure [13]. While classical turbulence models commonly assume isotropic and homogeneous conditions, actual atmospheric environments-especially near the ground and in urban settings exhibit substantial anisotropy caused by wind shear, temperature stratification, and layered atmospheric flows [14-16]. Since the strength of anisotropic turbulence varies in slant-path propagation along ground-to-air, air-to-ground, and cross-terrain optical communications, the propagation evolution and stability of structured beams depend on altitude dependent turbulent regimes over these paths. Altitude-dependent turbulence regimes along these pathways lead to cumulative distortions that significantly influence the evolution and stability of structured beams. Despite the increasing interest in vector beam propagation, the combined effects of anisotropic turbulence and slant-path geometry are still inadequately explored. Most existing theoretical approaches are either centered on scalar optical fields or rely on oversimplified isotropic turbulence models, which limits their ability to accurately describe polarization-spatial coupling, coherence loss, and beam wandering effects in vectorial optical fields under realistic atmospheric conditions. Consequently, a detailed and systematic investigation is required to reliably evaluate the robustness of structured vector beams and to assess their viability for long-range free-space optical applications [17-20].

In the following, the impact of atmospheric turbulence has been investigated for vector vortex beams to study their characteristics along vertical, slanted and horizontal paths. Our model integrates anisotropic refractive-index power spectra with altitude-dependent turbulence intensity profiles, effectively representing essential characteristics of actual atmospheric conditions experienced in ground-to-air and cross-terrain wireless optical communications. The characterisation of C- and V- point singularities in LG and circular Airy vortex (CAV) beams is conducted by examining alterations in their intensity profiles and spatial polarization structure as turbulence strength alters, offering insights into turbulence-induced distortion and depolarization effects. Furthermore, scintillation index (SI) is analyzed as a function of propagation distance to further examine the impacts of turbulence on LG and Airy vector beams that support various polarization singularities. The results delineate both the advantages and limitations of vector beams in complex turbulence environments. These findings offer crucial design principles for next-generation free-space optical communication systems and sophisticated imaging technologies functioning under realistic atmospheric circumstances.

## 2. Theoretical framework

The propagation of vector vortex beam launched at the plane $z = 0$ and propagating along a slant-path in atmosphere with anisotropic turbulence is shown schematically in fig 1.

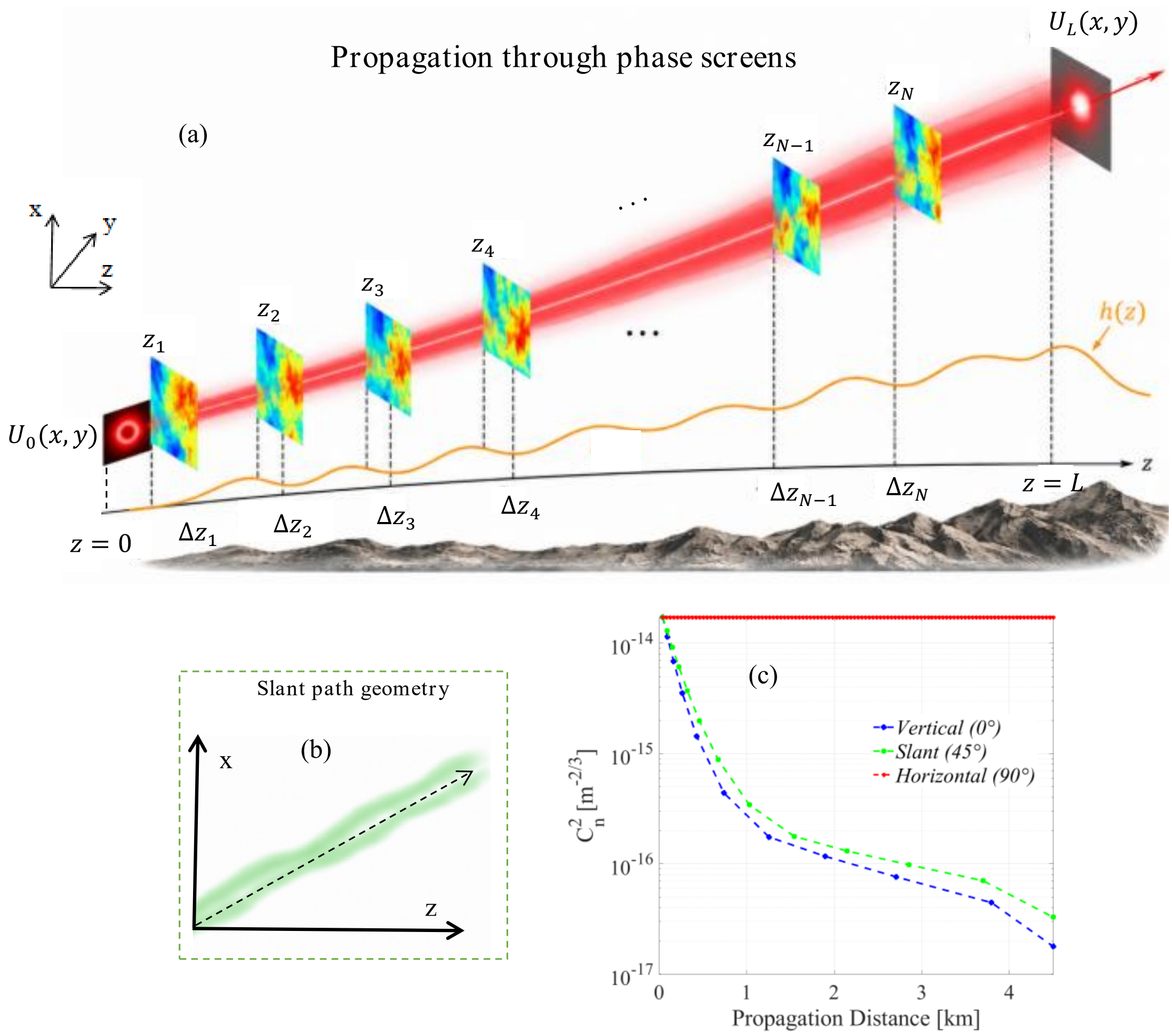

Fig. 1: (a) Illustration of vector vortex beam $U_0(x, y)$ propagation through atmospheric turbulence along a slant path using the multiple phase-screen method. The green envelope indicates beam spreading, scintillation, and wavefront distortion, while the blue dashed curve $h(z)$ represents altitude variation. The output field $U_L(x, y)$ shows the distorted beam intensity and polarization after propagation. (b) Geometrical representation of the slant-path propagation, where $\theta$ is the zenith/slant angle between the propagation axis and the reference vertical direction. (c) Variation of the refractive-index structure parameter $C_n^2$ with propagation distance for vertical($\theta = 0^{\circ}$), (b) slanted($\theta = 45^{\circ}$), and (c) horizontal paths ($\theta = 90^{\circ}$) paths, highlighting the dependence of turbulence strength on propagation geometry.

The optical fields at the source plane of vector LG [19] and CAV beams are given respectively by

$$U_{LG}(r,\phi) = \left[\left(\frac{r}{w_0}\right)^{|m|} e^{im\phi} e^{i\delta_1} \hat{e}_R + \left(\frac{r}{w_0}\right)^{|n|} e^{in\phi} e^{i\delta_2} \hat{e}_L\right] \times exp\left(-\frac{r^2}{w_0^2}\right), \tag{1}$$

and

$$U_{CAV}(r,\phi) = \left[e^{im\phi} e^{i\delta_1} \hat{e}_R + e^{in\phi} e^{i\delta_2} \hat{e}_L\right] \times airy\left(\frac{R_0 - r}{w_0}\right) exp\left(b\frac{R_0 - r}{w_0}\right), \tag{2}$$

where $r$ is the radial co-ordinate, $\phi$ is the azimuthal angle, the Airy function is denoted as $airy(.)$, while the initial radius $R_0$ of the ring is associated with the spatial extent of the beam. The radius of the beam waist is represented with $w_0$, the parameter $b$ is the truncation factor $(0 < b < 1)$, and $m$ & $n$ are integers that denote the topological charge of the independent scaler vortex beam. $\hat{e}_R$ and $\hat{e}_L$ are the basis vectors for right-circularly and left-circularly polarized beams respectively.

For vector beams characterized by azimuthal indices $m = -n$, the polarization structure corresponds to a V-point singularity. In contrast, when $m \neq n$, the polarization state is elliptical and the field supports C-point polarization singularity.

**Propagation model in anisotropic non-Kolmogorov turbulence**

To model atmospheric refractive-index fluctuations, we adopt a generalized anisotropic non-Kolmogorov power spectral density (PSD), which allows for both variable spectral slope and directional anisotropy. Such three-dimensional refractive-index spectrum is expressed as [20, 21]

$$\Phi_n(\kappa_x, \kappa_y, \kappa_z; \alpha) = A_\alpha \tilde{C}_n^2 (\mu_x^2 \kappa_x^2 + \mu_y^2 \kappa_y^2 + \kappa_z^2)^{-\alpha/2}, \tag{3}$$

where $\kappa_x, \kappa_y$, and $\kappa_z$ are the spatial wavenumbers in the three cartesian directions. The parameters $\mu_x$ and $\mu_y$ characterize the degree of anisotropy in the transverse directions, while α is the spectral slope satisfying $3 < \alpha < 4$. The quantity $\tilde{C}_n^2$ is the generalized refractive-index structure parameter with units of $m^{3-\alpha}$, and the normalization constant $A_\alpha$ is defined as

$$A_\alpha = \frac{\Gamma(\alpha - 1)}{4\pi^2} \cos\left(\frac{\pi\alpha}{2}\right). \tag{4}$$

Here, $\Gamma(.)$ denotes the gamma function. When $\mu_x = \mu_y$ and $\alpha = 11/3$, Eq. (3) reduces to the conventional isotropic Kolmogorov turbulence spectrum.

**Phase-screen approximation**

For numerical modelling of beam propagation, we employ the split-step phase-screen method, which is well suited for long-distance atmospheric links [22]. Under the Markov approximation, refractive-index fluctuations are assumed to be delta-correlated along the propagation direction [23]. Consequently, the turbulence-induced phase perturbations are represented by a sequence of statistically independent thin phase screens separated by free-space propagation segments.

Within this framework, the longitudinal wave-number component $\kappa_z$ is set to zero, since the phase screens act only in the transverse plane. Equation (3) then reduces to an effective two-dimensional phase PSD given by

$$\Phi_\phi(\kappa_x, \kappa_{y,}, \kappa_z = 0; \alpha) = 2\pi k^2 \Delta z\, A_\alpha \tilde{C}_n^2 (\mu_x^2 \kappa_x^2 + \mu_y^2 \kappa_y^2)^{-\alpha/2}, \tag{5}$$

where $k = 2\pi/\lambda$ is the optical wave-number, $\lambda$ is the wavelength, and $\Delta z$ denotes the spacing between the random phase screens.

The random phase perturbation at each screen is synthesized in the Fourier domain as

$$\phi(x,y) = \mathcal{F}^{-1}\left\{\sqrt{\Phi_\phi(\kappa_x, \kappa_y; \alpha)}\left[\xi_1(\kappa_x, \kappa_y) + i\xi_2(\kappa_x, \kappa_y)\right]\right\}, \tag{6}$$

where $\xi_1$ and $\xi_2$ are independent zero-mean Gaussian random variables with unit variance, and $\mathcal{F}^{-1}$ denotes the inverse Fourier transform.

**Beam propagation between phase screens**

Between adjacent phase screens, the optical field propagates through free space and obeys the paraxial wave equation. In the spatial-frequency domain, the propagation over a distance $\Delta z$ is described by

$$U(x,y,z+\Delta z) = \mathcal{F}^{-1}\left\{\mathcal{F}\{U(x,y,z+\Delta z)\}exp\left(-i\frac{\kappa_x^2+\kappa_y^2}{2k}\Delta z\right)\right\}. \tag{7}$$

At each phase screen, the field is modified following

$$U(x,y,z) \to U(x,y,z)exp[i\phi(x,y)]. \tag{8}$$

The combined application of eqns. (6)-(8) constitutes a complete split-step numerical model for beam propagation through anisotropic non-Kolmogorov turbulence.

**Slant-path extension**

In a slant-path propagation scenario, the generalized refractive index structure parameter becomes a function of altitude, i.e., $\tilde{C}_n^2 = \tilde{C}_n^2(h)$ [24]. At ground level ($h = 0$), the turbulence strength is taken as $\tilde{C}_n^2 = 1.74 \times 10^{-14} m^{3-\alpha}$. As the beam propagates along the slant path, the altitude increases according to the geometrical relation $h = s\cos\theta$, where $s$ is the path length and $\theta$ is the zenith angle with respect to the vertical (fig. 1(b)). At each propagation step, the corresponding value of $\tilde{C}_n^2(h)$ is updated based on the current altitude.

To ensure that the turbulence-induced phase variance introduced by each phase screen remains within the validity of the split-step approximation, the longitudinal spacing between successive phase screens is not chosen to be constant. Instead, it is determined dynamically according to [25]

$$\Delta z_n = \left(\frac{c_1}{1.23\tilde{C}_n^2(z_{n-1})k^{7/6}}\right)^{6/11}, \tag{9}$$

where $k$ is the optical wave-number and $c_1$ is a constant controlling the strength of the phase fluctuations per screen. This relation ensures that each phase screen contributes an approximately uniform amount of turbulence-induced phase distortion, even when the turbulence strength varies with altitude.

**Stokes parameter formulation**

For a fully vectorial optical field propagating along $z$, the transverse electric field can be written as

$$\vec{E}(x,y,z) = \begin{bmatrix} E_x(x,y,z) \\ E_y(x,y,z) \end{bmatrix}. \tag{10}$$

The polarization state is described statistically using the Stokes parameters [26]:

$$S_0 = \langle |E_x|^2 + |E_y|^2 \rangle, \tag{11}$$

$$S_1 = \langle |E_x|^2 - |E_y|^2 \rangle, \tag{12}$$

$$S_2 = 2Re\langle E_x E_y^* \rangle, \tag{13}$$

$$S_3 = -2Im\langle E_x E_y^* \rangle, \tag{14}$$

where $\langle . \rangle$ denotes ensemble averaging.

The degree of polarization (DoP) is then defined as

$$DoP = \frac{\sqrt{S_1^2 + S_2^2 + S_3^2}}{S_0}. \tag{15}$$

For vector beams propagating through turbulence, random phase perturbations modify the field correlations $\langle E_x E_y^* \rangle$, leading to a reduction in DoP. This reduction is often called depolarization. To quantify the overall polarization preservation across the beam cross-section, the average degree of polarization (Avg DoP) is introduced as the spatially averaged value of the local DoP, providing a global measure of the beam's polarization purity after propagation. Mathematically, it is expressed as

$$\bar{P} = \frac{\iint P(x,y) S_0(x,y) dxdy}{\iint S_0(x,y) dxdy}, \tag{16}$$

where $P(x, y)$ is the local degree of polarization. A higher value of $\bar{P}$ indicates stronger polarization resilience against turbulence-induced depolarization.

**Simulation parameters**

For simulating the propagation dynamics of polarization singular CAV and LG beams in atmosphere with anisotropic turbulence, the computational aperture window of dimension $1.5\ m \times 1.5\ m$ was sampled on a $1024 \times 1024$ grid and free-space propagation between screens was performed with an angular-spectrum (FFT) propagator. A wavelength of 633 nm has been chosen because it lies within low-absorption atmospheric window and provides an appropriate balance between diffraction spreading and turbulence-induced phase distortions. In addition, this wavelength is readily available from stable, high-quality He-Ne sources and is well matched to spatial light modulators, so the same parameters used in the present simulations can be directly adopted in the planned experimental validation. The parameters for the CAV ($R_0 = 0.02, b = 0.1, w_0 = 1.5cm$) have been optimized for maximum robustness against turbulence induced distortion. Similarly, the beam width $w_0 = 1.5cm$ for the LG beam has been chosen to facilitate comparison with the CAV beam since the LG beam is a standard and extensively studied vector beam that serves as a reliable reference. The minimum number of phase screens necessary for various propagation paths has been taken as 11 for vertical, 13 for slanted ($\theta = 45^{\circ}$), and 107 for horizontal propagation as shown in fig. 1(c). Fewer phase screens are sufficient for vertical propagation because the cumulative turbulence effect is relatively weak. In contrast, the slanted path requires a larger number of screens owing to the increased altitude-dependent refractive-index fluctuations. The horizontal path demands the highest number of phase screens, with a minimum of 107 screens, since the beam propagates entirely through the near-ground region where refractive-index fluctuations remain continuously strong along the propagation direction [27].

## 3. Numerical Results and Discussions

The evolutions of polarization singular vector beams in turbulent environments are modelled using angular spectrum and phase screen methods with the slant path extension as discussed in section 2. The different characterization parameters have been observed over a propagation distance of 4.5 km. The distortions induced by turbulence in the transverse intensity and polarization structures have been quantified using spatially resolved Stokes parameters. Furthermore, we have investigated the evolution of the polarization state through their trajectories on the Poincaré sphere, which reveals turbulence-driven polarization scrambling.

### 3.1 Intensity and Polarization Distributions of Vector Beams

The propagation of vector beams through anisotropic turbulence along slant paths exhibit a strong dependence on the inclination angle of the optical link as depicted in fig. 2(b-d). The evolutions of the C-point (lemon) and V-point (azimuthal) singularities are presented for LG beams in the first two columns, while those for the Airy beams are shown in the third and fourth columns with (a1-a4) specifically representing the initial profiles of these beams. For vertical propagation (figs. 2 (b1-b4)), the transverse intensity and polarization structures of both LG and Airy beams remain largely preserved, whereas stronger slant angles ((c1-c4) and (d1-d4)) lead to pronounced beam broadening and polarization scrambling. As shown in Fig. 2, Airy beams demonstrate enhanced robustness of both intensity and polarization profiles owing to their inherent self-healing property, while LG beams suffer increased diffraction-induced degradation under strong slant-path conditions. The variations in beam characteristics observed along various inclined pathways can be attributed to altitude-dependent alterations in the refractive index. These changes affect the beam's modal structure symmetry and cause spatial variations in its polarization. Specifically, LG beams with azimuthal V-point singularities exhibit a greater susceptibility to diffraction spreading and cross-polarization coupling.

Conversely, Airy beams possess radially symmetric but spatially non-separable field structures that reduce turbulence coupling, and their C-point polarization singularities remain confined within the main lobe, offering greater resilience than both LG beams and V-point vector beams. A quantitative comparison of the polarization stability of C- and V-point vector beams is further carried out using the degree of polarization (DoP), which is discussed in detail in the next section.

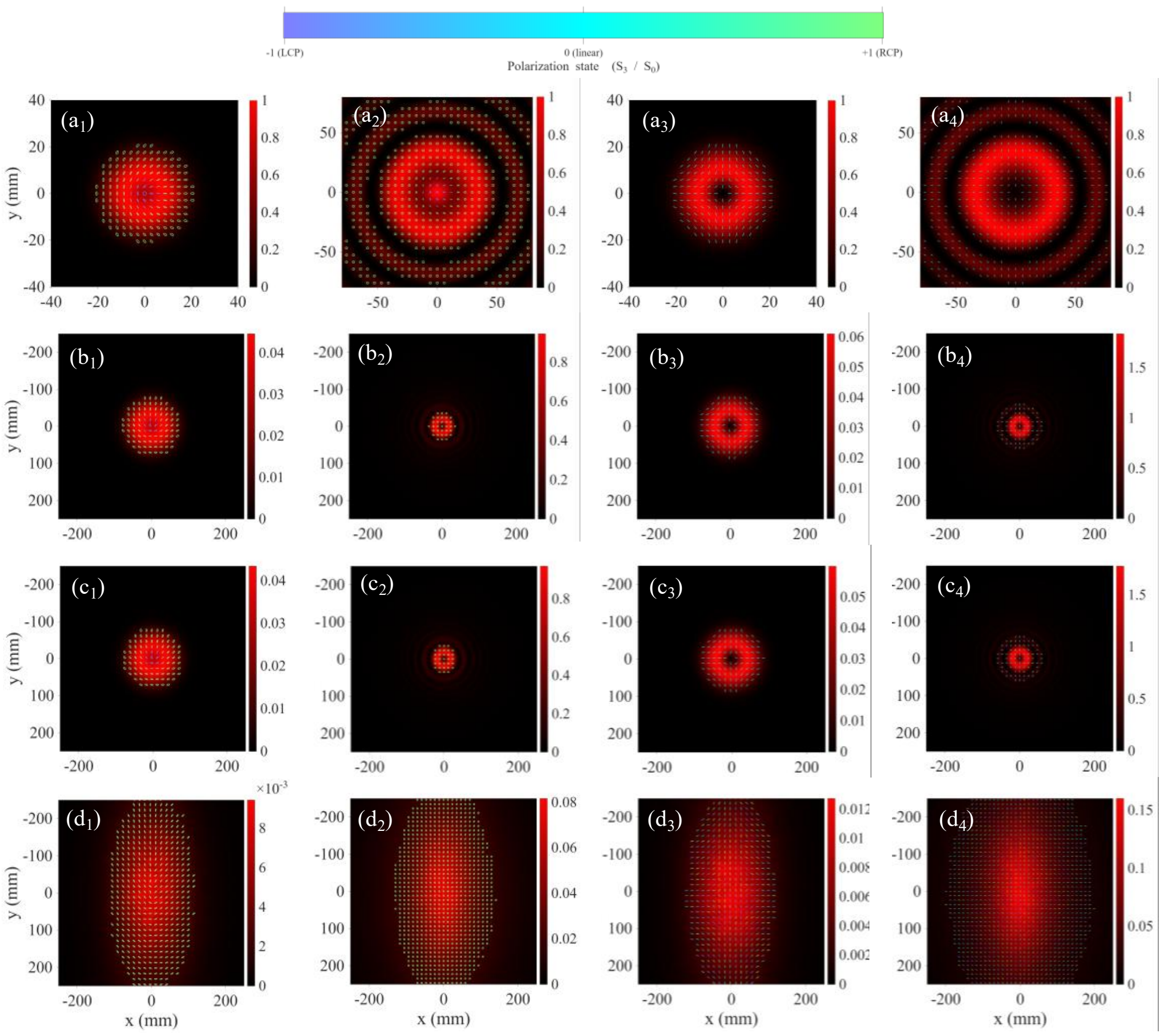

Figure 2: Evolutions of LG (1st and 3rd columns) and Airy (2nd and 4th columns) vector beams with C-type (lemon $m = 0$ and $n$ =1, $\delta_1 = 0, \delta_2 = 0$) and V-type (azimuthal $m = -1$ and $n$ =1, $\delta_1 = 0, \delta_2 = \pi$) polarization singularities in anisotropic atmospheric turbulence. The influence of turbulence on beam intensity and polarization over vertical, inclined, and horizontal transmission routes are illustrated in 2nd, 3rd and 4th rows, respectively. The accompanying horizontal colour bar represents the local polarization helicity, wherein blue colour corresponds to left-elliptical polarization (LEP), red colour denotes right-elliptical polarization (REP), and the central white region indicates linear polarization. The continuous transition of colours in the colour bar reflects the gradual spatial evolution of the polarization state.

**3.2 Scintillation index:**

Variation in transverse intensity profile of a beam caused by changes in the atmospheric refractive index characterizes the SI [23]. Here, we examine how the propagation distance affects the aperture-averaged scintillation [7] of vector beams traveling through an-isotropic ground-to-air routes as shown in fig. 3. Aperture averaging is performed at the observation plane using a circular receiving aperture with a diameter of 12 cm. These results assess the degradation of beam quality under practical conditions and analyze the turbulence tolerance across various types of vector beams.

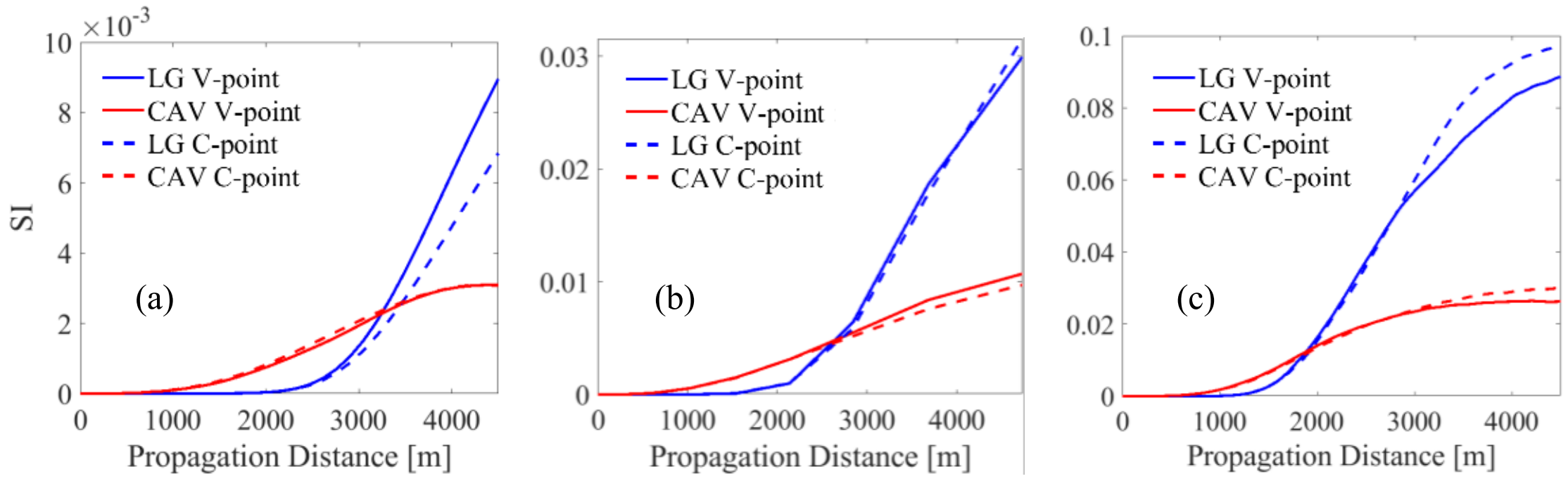

Figure 3: The variation of aperture averaged SI with propagation distance for LG and CAV beams with C-type and V-type singularities in an-isotropic turbulence condition along (a) vertical ($\theta = 0^{\circ}$), (b) slanted ($\theta = 45^{\circ}$), and (c) horizontal paths ($\theta = 90^{\circ}$).

Figure 3(a) illustrates that, during vertical propagation, the aperture-averaged SI increases gradually with propagation distance, in contrast to slanted or horizontal propagation, which experiences more turbulence fluctuations along the respective paths. LG beams exhibit lower SI for shorter ranges due to their symmetric ring-shaped intensity distribution so the variations at a receiving aperture location are low, however as the propagation distance increases LG beam undergoes greater fluctuations in speckle intensity owing to intensifying turbulence effect. While CAV beam exhibits lesser SI at $z = 4.5\ km$ because of its self-acceleration and focusing nature. Overall, V-point singularities display slightly larger values of SI than C-point configurations due to higher variation in angular polarization as well as their susceptibility to the perturbations caused by turbulence. For slant propagation (Fig. 3(b)), scintillation grows as the beam passes through several layers of altitude-dependent turbulence, resulting in cumulative phase distortions and greater irradiance variance. Here, we see the largest enhancement in SI for LG V-point beam under these conditions, highlighting the distinct combined effect of beam structure and polarization topology. Horizontal propagation (Fig. 3(c)) yields the highest overall SI values because continual exposure to near-ground an-isotropic turbulence enhances refractive index composite fluctuations and intensity instability. As shown in fig. 3(a-b), CAV beams show relatively less scintillation compared with other propagation geometries due to their self-healing property. In addition, the stability of C-point singularities is fairly strong compared with V-point singularities because the change of angular polarization in space for C-points remains smooth. Accordingly, the CAV beam structure and C-point topology, in combination, consistently yield the lowest scintillation and thus the most robust propagation in the presence of anisotropic atmospheric conditions.

### 3.3 Higher order Poincaré-Sphere (HOPS) Analysis of Polarization Evolution in Anisotropic Turbulence

In this section, we examine the evolution of polarization in various singularity points of vector beams in anisotropic turbulence along ground-to-air links. Figure 4 illustrates the HOPS representation of C-point singularity of LG and CAB using Stokes parameters. In the figure, the first column (a1-d1) depicts the polarization evolution of LG beams, and the second column (a2-d2) presents that of Airy beams. The first row of the figure represents the state of the beams at the input ($z = 0$) while 2$^{nd}$, 3$^{rd}$, and 4$^{th}$ rows respectively picture the polarization evolution in vertical, slanted, and horizontal transmissions at $z = 4.5$ km. The colour coding maps the local polarization states across the beam cross section, where red denotes right-handed elliptical polarization, blue indicates left-handed eliptical polarization, and green represents linear or near-linear polarization. LG beams exhibit spatial variation in both ellipticity and azimuthal angle, leading to pronounced polarization state diffusion and reduced average degree of polarization with increasing turbulence strength. Airy beams remain predominantly linearly polarized due to the absence of radial polarization dependence, resulting in more confined polarization trajectories and improved polarization stability.

For LG beams, significant spatial variation in both ellipticity and azimuthal angle leads to broad polarization distributions on the sphere even at the initial stage. As turbulence strength increases from vertical to horizontal propagation, these states exhibit enhanced diffusion and a noticeable reduction in the Avg DoP, reflecting stronger depolarization effects. In contrast, the Airy beams remain predominantly linear, as indicated by the concentration of green points along a confined trajectory, owing to the absence of radial polarization dependence. Although some spreading occurs under stronger turbulence, the Airy beams maintain more localized polarization states and higher Avg DoP values compared to LG beams, demonstrating improved polarization stability for C-point singularities.

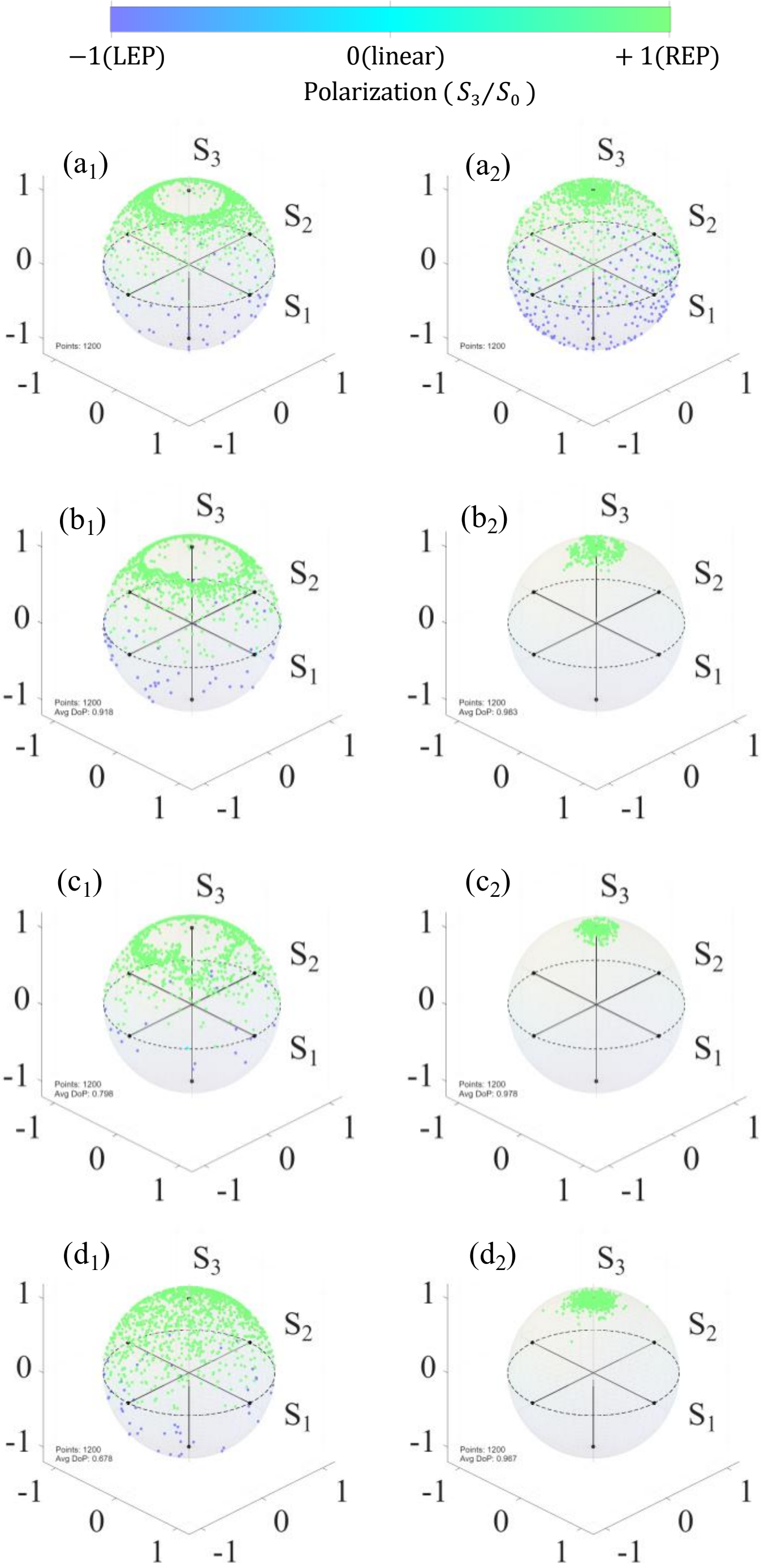

Figure 4: Polarization dynamics of C-type vector beams at the source plane (row 1) and after propagating through turbulence along vertical, slanted and horizontal propagation (row 2-4 respectively) represented on poincaré sphere. (a1-d1) correspond to LG beams and (a2-d2) to Airy beams. Avg DoP values are mentioned in the figures.

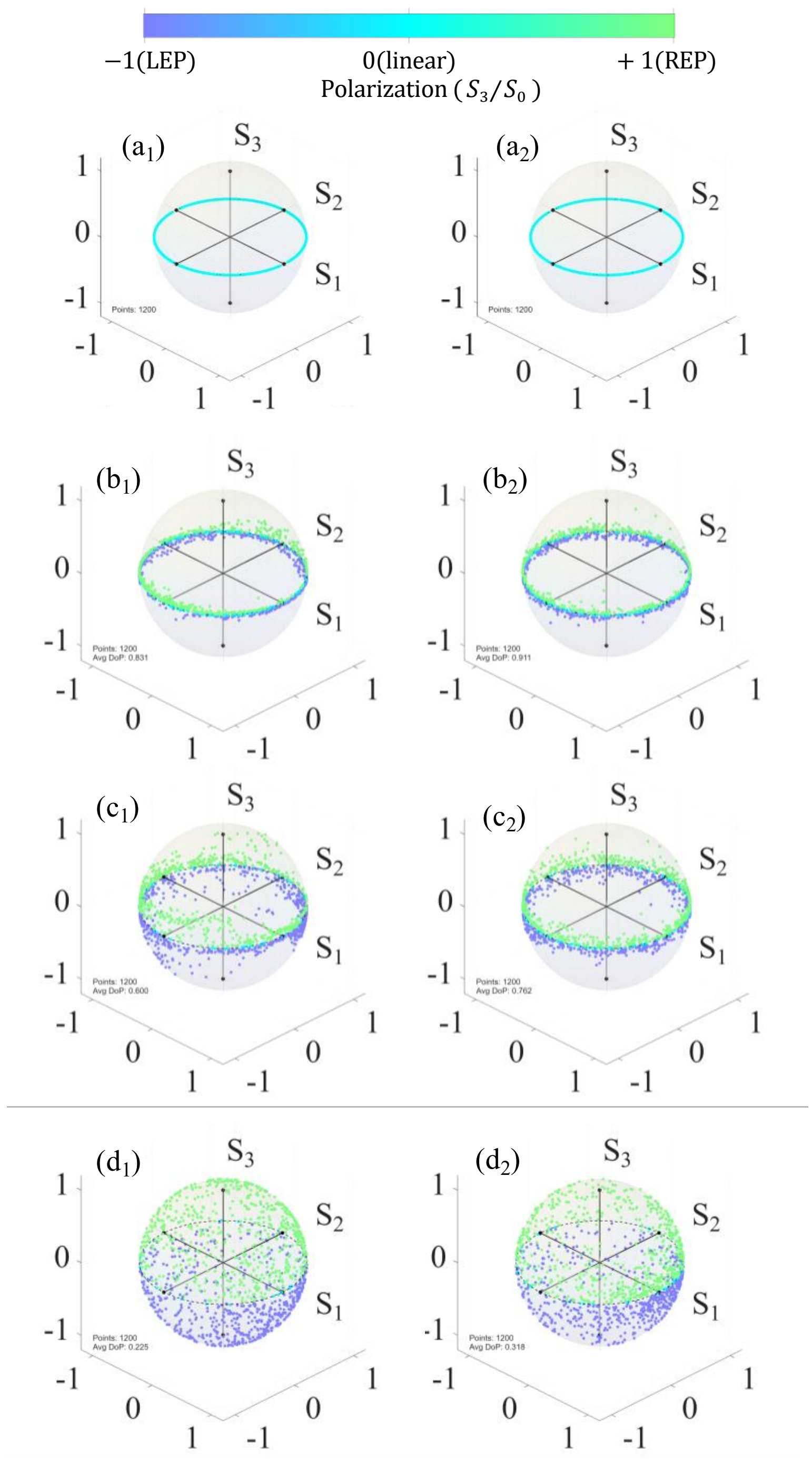

Figure 5: Polarization dynamics of V-type vector beams at the source plane (row 1) and after propagating through turbulence along vertical, slanted and horizontal propagation (row 2, 3 and 4 respectively) represented on Poincaré sphere. (a1-d1) correspond to LG beams and (a2-d2) to Airy beams. Avg DoP values are mentioned in the figures.

As shown in fig. 5, for V-point singularity, the polarization states remain concentrated near the equatorial region of the sphere, consistent with linear polarization, although turbulence-driven diffusion again increases from vertical to horizontal paths. In this case, LG beams experience stronger polarization spreading and lower Avg DoP compared to Airy beams.

Overall, C-point beams exhibit greater resilience to anisotropic turbulence than V-point beams due to smoother spatial polarization variation, with the Airy C-point configuration providing the highest polarization robustness under long-distance atmospheric propagation.

**CONCLUSION:**

We reported herein, a systematic exploration of structured vector beam propagation through an-isotropic atmospheric turbulence. Our focus mainly was on intensity fluctuations and polarization stability under realistic propagation geometries. The effects of an-isotropic turbulence on the propagation of a structured beam in vertical, slant-path and horizontal transmissions over a distance of 4.5 km have been investigated. The scintillation analysis demonstrated a progressive increase in intensity fluctuations from vertical to horizontal propagation configurations, highlighting the dominant role of near-ground an-isotropic turbulence. In all cases, Airy vector beams exhibited reduced scintillation compared to LG beams, confirming their enhanced resistance to turbulence-induced intensity modulation. The polarization behaviour of the propagated beams was quantified using the Stokes formalism and the degree of polarization, with the corresponding polarization state evolution visualized on the HOPS. The results revealed that anisotropic turbulence induces significant polarization diffusion and depolarization, particularly for horizontal and slant-path propagation. Compared to LG beams, Airy beams consistently maintained higher average degrees of polarization and more confined polarization trajectories, reflecting superior polarization coherence. Furthermore, a direct comparison between polarization singularities showed that C-point (lemon) beams are more resilient than V-point (azimuthal) beams, owing to smoother spatial polarization variation and reduced angular sensitivity to turbulence. Overall, the combination of Airy beam structure with C-point polarization provided the highest robustness against both intensity and polarization degradation in anisotropic atmospheric turbulence. These findings underscore the necessity of full vectorial modelling for accurate performance prediction of structured beams and provide valuable guidance for the design of polarization-sensitive free-space optical communication and sensing systems operating over long atmospheric paths.

**Data availability statement-**The data that support the findings of this study are available from the corresponding author upon reasonable request.

**Acknowledgement**
Sagar thanks the Ministry of Education (MoE), (formerly known as Ministry of Human Resource Development (MHRD)), Government of India, for scholarship assistance. This work is supported by the research grant CRG/2022/007736 from SERB (India).

**References:**

[1] S. Singh, S. K. Mishra, and A. K. Mishra, "Ring Pearcey vortex beam dynamics through atmospheric turbulence," J. Opt. Soc. Am. B, vol. 40, no. 9, 2023.

[2] W. Ren, H. Zhang, X. Mao, D. Xu, Z. Lu, W. Lin, S. Duan, and B. Liu, "Deep-learning-assisted orbital angular momentum mode recovery under atmospheric turbulence for free-space optical communication," J. Lightwave Technol., vol. 43, no. 18, pp. 8647, 2025. esn252280 KNaYv8vu

[3] A. Forbes, F. Nothlawala, and A. Vallés, "Progress in quantum structured light," Nat. Photonics, vol. 19, pp. 1291–1300, 2025.

[4] M. P. Edgar, D. S. Tasca, F. Izdebski, R. E. Warburton, J. Leach, M. Agnew, G. S. Buller, R. W. Boyd, and M. J. Padgett, "Imaging high-dimensional spatial entanglement with a camera," Nat. Commun., vol. 3, p. 984, 2012.

[5] F. Wang, W. Du, Q. Yuan, D. Liu, and S. Feng, "Wander of a Gaussian-beam wave propagating through Kolmogorov and non-Kolmogorov turbulence along laser-satellite communication uplink," Structure of Atmospheric Turbulence, vol. 13, no. 2, p. 162, 2022.

[6] Y. Zhang, C. Ke, Y. Xie, and Y. Zhang, "Experimental investigation of LG beam propagating in actual atmospheric turbulence," Results Phys., vol. 35, p. 105327, 2022.

[7] Sagar, M. Kala, S. Singh, S. K. Mishra, and A. K. Mishra, "Propagation dynamics of elliptical Airy vortex beam through turbulent atmosphere," J. Opt., vol. 28, p. 035607, 2026.

[8] S. Fu and C. Gao, “Influences of atmospheric turbulence effects on the orbital angular momentum spectra of vortex beams,” Photonics Res., vol. 4, no. 5, pp. B1–B4, 2016.

[9] P. Lochab, P. Senthilkumaran, and K. Khare, “Propagation of converging polarization singular beams through atmospheric turbulence,” Appl. Opt., vol. 58, no. 23, pp. 6335–6345, 2019.

[10] Q. Zhan, “Cylindrical vector beams: from mathematical concepts to applications,” Adv. Opt. Photonics, vol. 1, pp. 1–57, 2009.

[11] Z. Sun, J. Wang, Z. Li, T. Wu, Y. Yang, Y. Liu, and Y. Lu, “Stable propagation of Ince–Gaussian vector beams through atmospheric turbulence,” Opt. Commun., vol. 515, p. 128293, 2022.

[12] W. Cheng, J. W. Haus, and Q. Zhan, “Propagation of vector vortex beams through a turbulent atmosphere,” Opt. Express, vol. 17, no. 20, pp. 17829-17846, 2009.

[13] F. Wang, W. Du, Q. Yuan, D. Liu, and S. Feng, “A survey of structure of atmospheric turbulence in atmosphere and related turbulent effects,” Atmosphere, vol. 12, no. 12, p. 1608, 2021.

[14] M. Beason, L. Andrews, and R. Phillips, “Study on the effect of anisotropy on a propagating beam.”

[15] X. Xiao, D. G. Voelz, I. Toselli, and O. Korotkova, “Gaussian beam propagation in anisotropic turbulence along horizontal links: theory, simulation, and laboratory implementation,” Appl. Opt., vol. 55, no. 15, pp. 4079-4084, 2016.

[16] M. Cheng, L. Guo, J. Li, and Q. Huang, “Propagation properties of an optical vortex carried by a Bessel-Gaussian beam in anisotropic turbulence,” J. Opt. Soc. Am. A, vol. 33, no. 8, pp. 1442-1450, 2016.

[17] Y. Na and D. K. Ko, “Adaptive demodulation by deep-learning-based identification of fractional orbital angular momentum modes with structural distortion due to atmospheric turbulence,” Sci. Rep., vol. 11, p. 23505, 2021.

[18] P. G. Michailidis, A. Brimis, and D. G. Papazoglou, “Assessing the performance of free-space optical links utilizing structured light: the case of Tornado waves,” J. Opt., vol. 27, no. 12, p. 125602, 2025.

[19] R. Joshi, B. Komal, G. Arora, S. Kumar, and P. Senthilkumaran, “Superior self-healing properties of generic Poincaré beam,” J. Opt., vol. 27, p. 065604, 2025.

[20] L. Cui, B. Xue, and F. Zhou, “Generalized anisotropic turbulence spectra and applications in the optical waves’ propagation through anisotropic turbulence,” Opt. Express, vol. 23, no. 23, pp. 30088-30103, 2015.

[21] H. Wang, M. Guan, H. Xiao, H. Jiang, C. Zhang, and Y. Liu, “Quadratic vortex optical pin beam with OAM mode anti-degradation characteristic,” IEEE Photonics J., vol. 17, no. 2, p. 7900108, 2025.

[22] J. W. Goodman, *Introduction to Fourier Optics*, McGraw-Hill Press, 1996.

[23] P. Lochab, P. Senthilkumaran, and K. Khare, *OAM States of Light*, 1st ed. Bristol, U.K.: IOP Publishing, 2020.

[24] ITU-R, “On propagation data and prediction methods required for the design of space-to-earth and earth-to-space optical communication systems,” ITU-R Document 3J/31-E, Radio-Communication Study Group Meeting, p. 7, 2001.

[25] K. Yang, S. Tang, X. Yang, and R. Zhang, “Propagation characteristics of a ring Airy vortex beam in slant atmospheric turbulence,” J. Opt. Soc. Am. B, vol. 38, no. 5, pp. 1510-1517, 2021.

[26] D. H. Goldstein, *Polarized Light*. Boca Raton, FL: CRC Press, 2011.